\documentclass[twocolumn,showpacs,showkeys,preprintnumbers,
               amsmath,amssymb,floatfix,10pt]{revtex4}
\usepackage{graphicx}
\usepackage{subfigure}
\usepackage{amsmath}
\usepackage{amssymb}
\usepackage{wrapfig}
\usepackage{color}
\usepackage[english]{babel}
\usepackage[warning,math]{easyeqn}
\usepackage[thinlines]{easybmat}

\def\<{\langle}
\def\>{\rangle}
\begin{document}
%
\preprint{}
%
\title{Functional modes of proteins are among the most robust ones}
\author{S. Nicolay}
\author{Y.--H. Sanejouand}
\affiliation{Laboratoire de Physique,
Ecole Normale Sup\'erieure, 46 all\'ees d Italie,   
69364 Lyon Cedex 07, France.}
%
\begin{abstract} 

It is shown that a small subset of modes which are likely
to be involved in protein functional motions of large amplitude can be determined
by retaining the most robust normal modes obtained using different protein models.
This result
should prove helpful in the context of several applications proposed recently, like 
for solving difficult molecular replacement 
problems or for fitting atomic structures into
low-resolution electron density maps.
Moreover, it may also pave the way for
the development of methods allowing to predict such motions accurately. 

\end{abstract} 
%
%
%
%
%
%
\maketitle
%
%
In the case of two-domain proteins, it is well known
that a few low-frequency normal modes can provide a fair description
of their large amplitude motion upon ligand binding\cite{Karplus:76,
Harrison:84,Brooks:85}. More recently, it has been shown that this is also
true for proteins with more complex architectures\cite{Marques:95,Perahia:95,Ma:05}, 
as long as their
functional motion is a collective one, {\it i.e.} if it concerns large
parts of the structure\cite{Tama:01,Delarue:02,Gerstein:02}. For instance,
a single low-frequency mode of the
T form of hemoglobin is enough to describe accurately its conformational change
upon oxygen binding\cite{Perahia:95}.
 
This result has been successfully applied for exploiting fiber diffraction
data\cite{Tirion:95}, solving difficult molecular replacement 
problems\cite{Elnemo1,Elnemo2,Delarue:04}, or fitting atomic structures into
low-resolution electron density maps\cite{Delarue:04,Tama:04,Hinsen:05}.
The principle of these applications is to perturb a known structure
along its low-frequency modes so as to get a deformed structure that is
consistent with low-resolution biophysical data, which are obtained after the protein
has undergone some large amplitude conformational change.
It was also shown
that when variations of a few key distances are known, through spectroscopic
measurements for instance, it is possible, using linear response theory,
to identify which modes are the most involved 
in the conformational change\cite{Kidera:05,Brooks:05}.
However, if such experimental data are missing, it is difficult to guess
which low-frequency modes are the functional ones. 
Hereafter, we show that they are among the most robust
ones, {\it i.e.} among the most conserved modes when different descriptions
of a given protein are considered. The robustness of the functional modes 
was recognized when it was shown that they can be obtained\cite{Tama:01,
Delarue:02,Gerstein:02} with simple protein descriptions, like Elastic
Network (EN) models\cite{Tirion:96,Bahar:97,Hinsen:98}. Herein, this property
is used so as to identify them.

First, standard normal modes were calculated for a set of five proteins
of different sizes and architectures after preliminary energy-minimization.
The CHARMM program\cite{Charmm} was used, with the EEF1.1 implicit solvent model
and the corresponding electrostatic and non-bonded options\cite{EEF11},
as done in recent studies performed at this level of detail\cite{Cui:04}.
Then, for each energy-minimized structure, low-frequency normal modes were calculated
with the all-atom EN model proposed by M. Tirion\cite{Tirion:96},
where the standard, many-parameters, empirical energy function $E_p$ used in
programs like CHARMM is replaced by:
\begin{equation}
   \label{ENM}
  E_p=\sum_{d_{ij}^0 < R_c} C(d_{ij}-d_{ij}^0)^2 
\end{equation}
where $d_{ij}$ is the distance between atoms $i$ and $j$, 
$d_{ij}^0$ being their distance in the 
studied structure.
The strength of the potential $C$ is a
constant assumed to be the same for all interactings pairs. It
is required only in order to define
energy (and frequency) units. As done in previous studies\cite{Elnemo1},
$R_c$, the cut-off parameter, is set to 5{\AA}.
\begin{figure}[t]
\centering
\includegraphics[width=8. truecm,angle=270,clip]{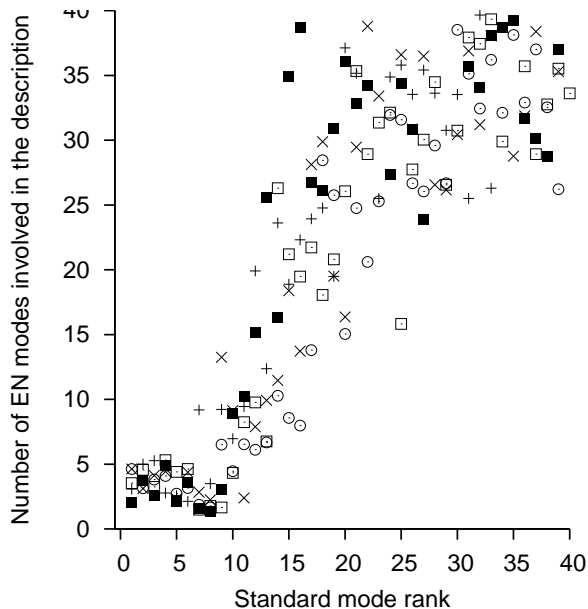}
\caption{\small \label{comp} 
Effective number of EN normal modes involved in the description of each standard mode
of five proteins. Cross: Lysozyme T4 (pdb code 178l).
Plus: Adenylate Kinase (4ake). Open square: Glutamin Binding Protein (1ggg).
Filled square: LAO Binding Protein (2lao). 
Open circle: DNA Polymerase $\beta$ (1bpx).
Modes are ranked according to increasing frequencies.
Modes ranked 1 to 6 correspond to the six, zero-frequency,  
rigid-body translations and rotations of each protein.}
\end{figure}

In order to compare both sets of normal modes, 
$n^{eff}_i$, the effective number of EN modes involved in the description
of standard 
mode $i$, 
is calculated as follows\cite{Local}:
\begin{equation}
\label{neff}
n^{eff}_i = \exp(- \sum^n \alpha I_{ij}^2 \ln(\alpha I_{ij}^2))
  \notag
\end{equation}
where 
$n$ is the number of EN modes taken into account ($n$=100 herein),
$I_{ij}$ being the scalar product between standard mode $i$ and
EN mode $j$.
The normalization factor
$\alpha$ is such that: $\sum \alpha I_{ij}^2 = 1$.
Thus, $n^{eff}_i$ gives the effective number of non-zero 
(normalized) $I_{ij}^2$.
It ranges from 1 to $n$. 
As shown in Fig. \ref{comp}, for each protein considered, several 
of its standard normal modes can be described accurately with less than 5-6 EN modes. 
Moreover, all these robust modes have low rankings, namely, below \#15.

Next, two other EN models were considered. In both cases, as often done,
\cite{Bahar:97,Hinsen:98,Tama:01,Bahar:01,Delarue:02,Gerstein:02}
only C$_\alpha$ atoms are kept. In the first model, as proposed by M. Tirion
[see Eq. (\ref{ENM})], pairs of interacting neighbors are determined according to
a distance-cutoff criterion, namely, $R_c=12$\AA. 
With such a criterion, for Adenylate Kinase, $n_c$,
the average number of interacting neighbors 
per C$_\alpha$ atom,
is 25 $\pm$ 7, ranging from 10 to 42,
as a function of the degree of burial of the amino-acid in the protein interior.
Note that $R_c$ can not be set to a value lower than 8-10\AA, a limit which depends
upon the structure considered. 
Otherwise, the number of zero-frequency modes becomes larger than six, 
as a consequence of the splitting of the elastic network 
into several independant ones.

The second model was designed so as to keep $n_c$ 
as constant as possible from
one amino-acid to the other. To do so, we use the following algorithm. First, all pairs
of C$_\alpha$ atoms are sorted, according to their distance. Then, 
starting from the pair separated by the largest distance,
they are removed one after the other, 
unless one atom of the pair has
already $n_c$ neighbors. With this algorithm, 
setting $n_c=10$, the average distance
between pairs of interacting neighbors is 6.2 $\pm$ 1.8\AA, ranging from 3.0 to 10.8\AA.
Note that in the case of Adenylate Kinase $n_c$ can be set to a value as low as 7 
without splitting the network
into independant ones.
\begin{figure}[t!]
\centering
\includegraphics[angle=270,width=8. truecm,clip]{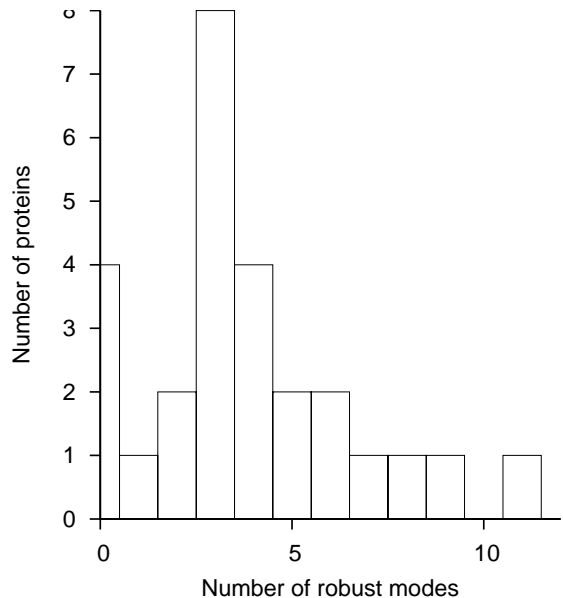}
\caption{\small \label{robust} 
Number of robust normal modes found by comparing modes obtained with
different protein models. For a first set of five proteins, standard modes
were compared to modes obtained with an all-atom EN model. For a second set of twenty-two
proteins, modes obtained with two different C$_\alpha$-EN model were compared.
Modes are considered to be robust when they can be described accurately with at most six modes
obtained with another protein model.}
\end{figure}

As done above, normal modes obtained with both EN models were compared,
seeking for robust ones, using a set of twenty-two proteins considered in previous
studies performed with the distance-cutoff criterion\cite{Tama:01,Delarue:02,Elnemo1}. 
Like in the case of all-atom models, modes are considered to be robust whenever 
$n^{eff}_i \leq 6$. 

Statistics of the number of robust modes found for all
studied proteins are shown in Fig. \ref{robust}
(zero-frequency modes are not taken into account). In most cases, 
the number of robust modes is four or less. In only three cases,  
it is larger than seven. Interestingly, 
the DNA polymerase of bacteriophage RB69 (pdb code 1ih7),
which is the protein of our dataset with the largest
number of robust modes (eleven),  
has a quite complex architecture, with three well-known structural domains.
It is also among the largest cases considered herein (897 amino-acids).

In four cases, no robust mode is found. Interestingly, the known conformational
change of these proteins, namely, Tyrosine Phosphatase, Triose Phosphate Isomerase,
Che Y, and HIV-1 protease (pdb codes are 1yts, 3tim, 3chy, 1hhp, respectively),
is a small amplitude one,
with a C$_\alpha$
root-mean-square displacement (r.m.s.d.) 
of 1.5{\AA} at most.

Then, it was checked that robust modes yield accurate descriptions
of protein functional motions. To do so, $Q_d$, the quality of the motion
description is calculated as follows\cite{Perahia:95,Delarue:02}:
\begin{equation}
Q_d = 100 \sum_{i=1}^{n} I_{id}^2
\notag
\end{equation}
where $n$ is the number of modes taken into account in the description and 
$I_{id}$ is the scalar product between mode $i$ and 
the direction of the conformational change observed by crystallographers.
Note that $Q_d=100$\% when all 
modes are included in the description, since they form a complete basis set\cite{Goldstein:50}.

\begin{figure}[t!]
\vskip -2cm
\includegraphics[width=8. truecm,clip]{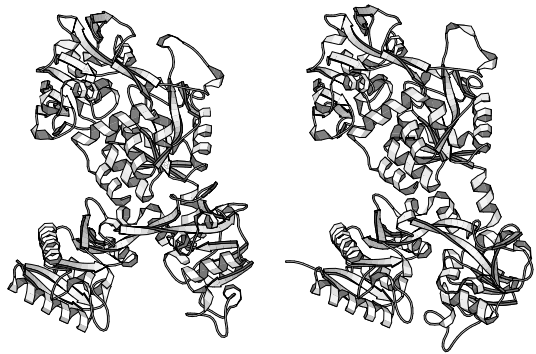}
\vskip -2cm
\caption{\small \label{lactof} The conformational change of Lactoferrin
upon ligand binding. Left: apo (or "open") state (pdb code 1cb6).
Right: holo (or "closed") state (1lfg). In the latter case, the iron ligands are not shown.
Drawn with Molscript\cite{Molscript}.}
\end{figure}

\begin{figure}[t!]
\centering
\includegraphics[angle=270,width=8. truecm,clip]{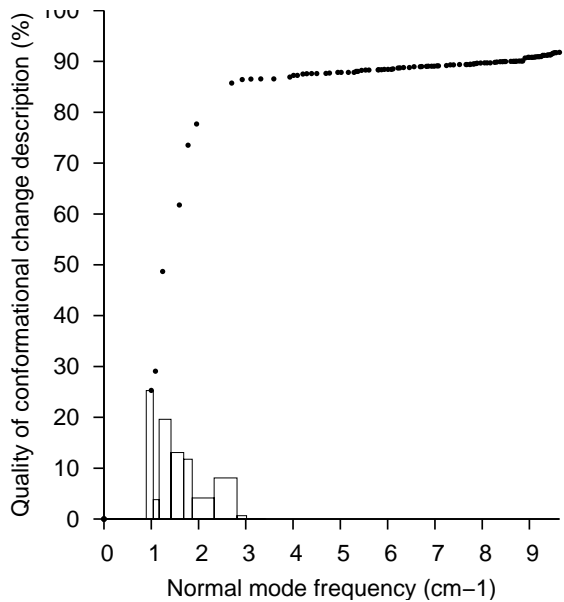}
\caption{\small \label{dconf} Quality of the description of the closure
motion of Lactoferrin upon ligand binding,
as a function of the number of low-frequency normal modes (black points) considered. 
Boxes: contribution of each robust mode to the description.} 
\end{figure}

\begin{figure}[t!]
\centering
\includegraphics[angle=270,width=8. truecm,clip]{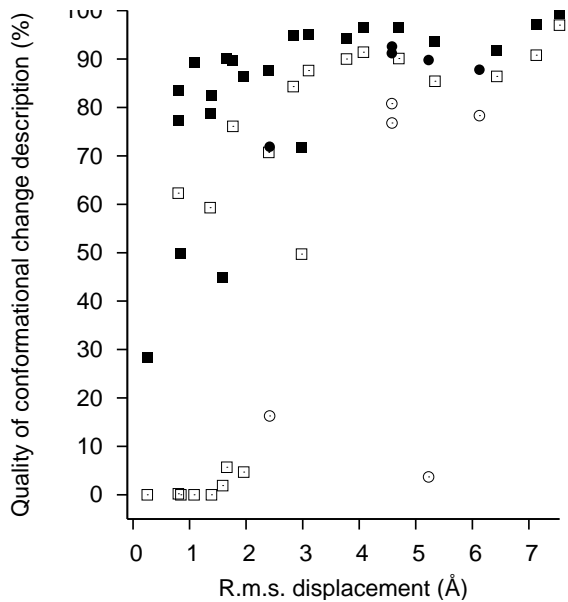}
\caption{\small \label{qual} Quality of the description of protein functional
motions with 100 low-frequency modes (filled symbols) or with only the robust ones
(open symbols), as a function of the amplitude of the motion.
Five proteins were
studied at the all-atom level (circles) and the other ones at the
amino-acid level (squares).
}
\end{figure}

In Fig. \ref{lactof}, the conformational change of lactoferrin is shown. 
It can be described accurately ($Q_d$ over 85\%) as a linear combination
of the seven lowest-frequency modes of the "open" form
(see Fig. \ref{dconf}). Interestingly, all seven modes are found to be robust.
In Fig. \ref{qual}, $Q_d$ is given 
as a function of the amplitude of the 
functional motion of each protein considered
when $n=100$ normal modes 
or when only the robust ones
are taken into account in the description. 
For most proteins with small amplitude motions, {\it i.e.} of less than 2-3{\AA} of r.m.s.d.,
robust modes fail to capture any information about the nature of the known 
conformational change, while in several cases some information is indeed present
in the normal modes. For instance, as mentioned above, for HIV-1 protease, 
no robust mode is found, although a single EN mode is enough for describing
50\% of its conformational change upon ligand binding\cite{Tama:01}. If two
other EN modes are added to the description, $Q_d$ can reach a value of 77\%
(with $n$=100, $Q_d$=89\%). 

On the other hand, when considering proteins with large 
amplitude motions, 
the description of
the conformational change with robust modes is almost as accurate 
($Q_d$ over 75\%)
as when $n=100$ normal modes are taken into account.
The only counter exemple is Adenylate Kinase, whose r.m.s.d. upon
ligand binding is 5.3{\AA} (the corresponding pdb codes of the open and closed
crystallographic structures are 4ake and 1ake). As a matter of fact, when standard normal
modes of Adenylate Kinase are compared to all-atom EN ones, only a single 
robust mode is found (see Fig. \ref{comp}), and it is not 
involved in the conformational change ($Q_d$=4\%). However,
using C$_\alpha$-EN models,
six robust modes are found and they allow for an almost perfect description
of the conformational change ($Q_d$=91\%). 

Of course, when using all atom models, more robust modes can be obtained by raising
the robustness criterion. In the case of Adenylate Kinase, if a given mode 
is said robust whenever $n_i^{eff} \le 10$, then five robust modes
are found. However, it is still not enough 
($Q_d$=73\%)
for describing its conformational change as well as with robust modes 
obtained using C$_\alpha$-EN models. Raising the robustness criterion
so as to obtain six robust modes does not change significantly
the quality of the description ($Q_d$=77\%).
As a matter of fact, robust modes obtained using all-atom models always yield poorer description
of protein functional motions than simpler models, in which only $C_\alpha$ atoms
are kept (open circles are below open squares in Fig. \ref{qual}).
This is mainly due to the fact that standard normal mode analysis requires
a preliminary energy-minimization, during which the structure is significantly
distorted, while normal mode analysis of EN models does not, as illustrated by the
case of DNA polymerase $\beta$. For this protein, when the C$_\alpha$-EN model is built 
using the crystal structure (pdb code 1bpx), seven robust modes are found,
which are able to describe accurately 
($Q_d$=84\%) 
the conformational change upon nucleotide binding (pdb code 1bpy).
However, when it is built using
the energy-minimized structure, only three robust modes are found, which are not
able to describe the conformational change ($Q_d$=21\%) much better than
the three ones obtained using all-atom models ($Q_d$=16\%). In that
case, the distortion during the energy-minimization process is unusually large
(r.m.s.d.=2.5\AA), probably as a consequence of the removal of the large ligand,
namely, a sixteen base pair DNA 
(1bpx is the structure of a binary complex while 1bpy is the structure
of a ternary complex), prior to the calculation. Even though the amplitude
of the distortion is almost as large as the amplitude of the functional motion
itself (r.m.s.d.=2.8\AA), the above result is not straightforward, since
the distortion does not occur along the direction of the conformational change.
Indeed, with respect to the energy-minimized structure, the amplitude of the
functional motion remains large (r.m.s.d.=2.4\AA).

In the present study, modes obtained with standard all-atom, many parameters, protein models
were compared to those obtained with elastic network models, as 
proposed by M. Tirion\cite{Tirion:96}.
For most protein cases, several robust modes are found, confirming results
obtained previously\cite{Tirion:96,Bahar:97,Hinsen:98,Tama:01}, namely, that the
lowest-frequency modes are little sensitive to details in the protein description.
Since such EN models rely on a distance-cutoff criterion
for defining atomic interactions, this can be explained in two
different ways. First, these modes may capture informations about 
the protein mass distribution in space. 
Second, they may capture informations about the rigidity of the protein in the vicinity
of each amino-acid residue. Indeed, with a distance-cutoff criterion, amino-acids
in the protein interior are more rigid (more neighbors) than those on the surface 
(less neighbors). So, we designed
a novel C$_\alpha$-EN model
whose main {\it raison d'\^etre} was to decide between these two possibilities.
In this model, each C$_\alpha$ atom has a given number of
interacting neighbors and
rigidity is fairly constant from one point of a protein to another.
When modes obtained with this model are compared to
those obtained with a 
C$_\alpha$-EN model based on the distance-cutoff criterion, the same 
robust modes are found.
This means that they are also not sensitive to the distribution of rigidity
in the protein.  

Moreover, we have shown that these robust modes are likely
to be involved in protein functional motions, at least when
the functional motion is a large amplitude one (r.m.s.d. $\ge$ 2-3\AA). This result
should prove helpful in the context of applications like those mentioned in the Introduction,
since they all concern large amplitude conformational changes
\cite{Elnemo1,Elnemo2,Delarue:04,Tama:04,Hinsen:05}. 

This result could also pave the way for
the development of methods allowing to predict such motions accurately, 
{\it i.e.} to predict their amplitude,
since 
exploring a subspace of small dimensionality (three or
four in most cases considered) 
should be enough for finding conformations close to functional ones.
Interestingly, seeking for robust modes could also indicate whether a given protein
can exhibit large amplitude functional motions or not. Indeed, the functional motions
of the four proteins with no robust mode are small
amplitude ones.


\begin{thebibliography}{27}
\expandafter\ifx\csname natexlab\endcsname\relax\def\natexlab#1{#1}\fi
\expandafter\ifx\csname bibnamefont\endcsname\relax
  \def\bibnamefont#1{#1}\fi
\expandafter\ifx\csname bibfnamefont\endcsname\relax
  \def\bibfnamefont#1{#1}\fi
\expandafter\ifx\csname citenamefont\endcsname\relax
  \def\citenamefont#1{#1}\fi
\expandafter\ifx\csname url\endcsname\relax
  \def\url#1{\texttt{#1}}\fi
\expandafter\ifx\csname urlprefix\endcsname\relax\def\urlprefix{URL }\fi
\providecommand{\bibinfo}[2]{#2}
\providecommand{\eprint}[2][]{\url{#2}}

\bibitem[{\citenamefont{McCammon et~al.}(1976)\citenamefont{McCammon, Gelin,
  Karplus, and Wolynes}}]{Karplus:76}
\bibinfo{author}{\bibfnamefont{J.~A.} \bibnamefont{McCammon}},
  \bibinfo{author}{\bibfnamefont{B.~R.} \bibnamefont{Gelin}},
  \bibinfo{author}{\bibfnamefont{M.}~\bibnamefont{Karplus}}, \bibnamefont{and}
  \bibinfo{author}{\bibfnamefont{P.}~\bibnamefont{Wolynes}},
  \bibinfo{journal}{Nature} \textbf{\bibinfo{volume}{262}},
  \bibinfo{pages}{325} (\bibinfo{year}{1976}).

\bibitem[{\citenamefont{Harrison}(1984)}]{Harrison:84}
\bibinfo{author}{\bibfnamefont{W.}~\bibnamefont{Harrison}},
  \bibinfo{journal}{Biopolymers} \textbf{\bibinfo{volume}{23}},
  \bibinfo{pages}{2943} (\bibinfo{year}{1984}).

\bibitem[{\citenamefont{Brooks and Karplus}(1985)}]{Brooks:85}
\bibinfo{author}{\bibfnamefont{B.~R.} \bibnamefont{Brooks}} \bibnamefont{and}
  \bibinfo{author}{\bibfnamefont{M.}~\bibnamefont{Karplus}},
  \bibinfo{journal}{Proc. Natl. Acad. Sci. USA} \textbf{\bibinfo{volume}{82}},
  \bibinfo{pages}{4995} (\bibinfo{year}{1985}).

\bibitem[{\citenamefont{Marques and Sanejouand}(1995)}]{Marques:95}
\bibinfo{author}{\bibfnamefont{O.}~\bibnamefont{Marques}} \bibnamefont{and}
  \bibinfo{author}{\bibfnamefont{Y.-H.} \bibnamefont{Sanejouand}},
  \bibinfo{journal}{Proteins} \textbf{\bibinfo{volume}{23}},
  \bibinfo{pages}{557} (\bibinfo{year}{1995}).

\bibitem[{\citenamefont{Perahia and Mouawad}(1995)}]{Perahia:95}
\bibinfo{author}{\bibfnamefont{D.}~\bibnamefont{Perahia}} \bibnamefont{and}
  \bibinfo{author}{\bibfnamefont{L.}~\bibnamefont{Mouawad}},
  \bibinfo{journal}{Comput. Chem.} \textbf{\bibinfo{volume}{19}},
  \bibinfo{pages}{241} (\bibinfo{year}{1995}).

\bibitem[{\citenamefont{Ma}(2005)}]{Ma:05}
\bibinfo{author}{\bibfnamefont{J.}~\bibnamefont{Ma}},
  \bibinfo{journal}{Structure} \textbf{\bibinfo{volume}{13}},
  \bibinfo{pages}{373} (\bibinfo{year}{2005}).

\bibitem[{\citenamefont{Tama and Sanejouand}(2001)}]{Tama:01}
\bibinfo{author}{\bibfnamefont{F.}~\bibnamefont{Tama}} \bibnamefont{and}
  \bibinfo{author}{\bibfnamefont{Y.-H.} \bibnamefont{Sanejouand}},
  \bibinfo{journal}{Protein Engineering} \textbf{\bibinfo{volume}{14}},
  \bibinfo{pages}{1} (\bibinfo{year}{2001}).

\bibitem[{\citenamefont{Delarue and Sanejouand}(2002)}]{Delarue:02}
\bibinfo{author}{\bibfnamefont{M.}~\bibnamefont{Delarue}} \bibnamefont{and}
  \bibinfo{author}{\bibfnamefont{Y.-H.} \bibnamefont{Sanejouand}},
  \bibinfo{journal}{J. Mol. Biol.} \textbf{\bibinfo{volume}{320}},
  \bibinfo{pages}{1011} (\bibinfo{year}{2002}).

\bibitem[{\citenamefont{Krebs et~al.}(2002)\citenamefont{Krebs, Alexandrov,
  Wilson, Echols, Yu, and Gerstein}}]{Gerstein:02}
\bibinfo{author}{\bibfnamefont{W.~G.} \bibnamefont{Krebs}},
  \bibinfo{author}{\bibfnamefont{V.}~\bibnamefont{Alexandrov}},
  \bibinfo{author}{\bibfnamefont{C.~A.} \bibnamefont{Wilson}},
  \bibinfo{author}{\bibfnamefont{N.}~\bibnamefont{Echols}},
  \bibinfo{author}{\bibfnamefont{H.}~\bibnamefont{Yu}}, \bibnamefont{and}
  \bibinfo{author}{\bibfnamefont{M.}~\bibnamefont{Gerstein}},
  \bibinfo{journal}{Proteins} \textbf{\bibinfo{volume}{48}},
  \bibinfo{pages}{682} (\bibinfo{year}{2002}).

\bibitem[{\citenamefont{Tirion et~al.}(1995)\citenamefont{Tirion, ben Avraham,
  Lorenz, and Holmes}}]{Tirion:95}
\bibinfo{author}{\bibfnamefont{M.}~\bibnamefont{Tirion}},
  \bibinfo{author}{\bibfnamefont{D.}~\bibnamefont{ben Avraham}},
  \bibinfo{author}{\bibfnamefont{M.}~\bibnamefont{Lorenz}}, \bibnamefont{and}
  \bibinfo{author}{\bibfnamefont{K.}~\bibnamefont{Holmes}},
  \bibinfo{journal}{Biophys. J.} \textbf{\bibinfo{volume}{68}},
  \bibinfo{pages}{5} (\bibinfo{year}{1995}).

\bibitem[{\citenamefont{Suhre and Sanejouand}(2004{\natexlab{a}})}]{Elnemo1}
\bibinfo{author}{\bibfnamefont{K.}~\bibnamefont{Suhre}} \bibnamefont{and}
  \bibinfo{author}{\bibfnamefont{Y.-H.} \bibnamefont{Sanejouand}},
  \bibinfo{journal}{Act. Cryst. D} \textbf{\bibinfo{volume}{60}},
  \bibinfo{pages}{796} (\bibinfo{year}{2004}{\natexlab{a}}).

\bibitem[{\citenamefont{Suhre and Sanejouand}(2004{\natexlab{b}})}]{Elnemo2}
\bibinfo{author}{\bibfnamefont{K.}~\bibnamefont{Suhre}} \bibnamefont{and}
  \bibinfo{author}{\bibfnamefont{Y.-H.} \bibnamefont{Sanejouand}},
  \bibinfo{journal}{Nucl. Ac. Res.} \textbf{\bibinfo{volume}{32}},
  \bibinfo{pages}{W610} (\bibinfo{year}{2004}{\natexlab{b}}).

\bibitem[{\citenamefont{Delarue and Dumas}(2004)}]{Delarue:04}
\bibinfo{author}{\bibfnamefont{M.}~\bibnamefont{Delarue}} \bibnamefont{and}
  \bibinfo{author}{\bibfnamefont{P.}~\bibnamefont{Dumas}},
  \bibinfo{journal}{Proc. Natl. Acad. Sci. USA} \textbf{\bibinfo{volume}{101}},
  \bibinfo{pages}{6957} (\bibinfo{year}{2004}).

\bibitem[{\citenamefont{Tama et~al.}(2004)\citenamefont{Tama, Miyashita, and
  Brooks~III}}]{Tama:04}
\bibinfo{author}{\bibfnamefont{F.}~\bibnamefont{Tama}},
  \bibinfo{author}{\bibfnamefont{O.}~\bibnamefont{Miyashita}},
  \bibnamefont{and} \bibinfo{author}{\bibfnamefont{C.~L.}
  \bibnamefont{Brooks~III}}, \bibinfo{journal}{J. Mol. Biol.}
  \textbf{\bibinfo{volume}{337}}, \bibinfo{pages}{985} (\bibinfo{year}{2004}).

\bibitem[{\citenamefont{Hinsen et~al.}(2005)\citenamefont{Hinsen, Reuter,
  Navaza, Stokes, and Lacapere}}]{Hinsen:05}
\bibinfo{author}{\bibfnamefont{K.}~\bibnamefont{Hinsen}},
  \bibinfo{author}{\bibfnamefont{N.}~\bibnamefont{Reuter}},
  \bibinfo{author}{\bibfnamefont{J.}~\bibnamefont{Navaza}},
  \bibinfo{author}{\bibfnamefont{D.~L.} \bibnamefont{Stokes}},
  \bibnamefont{and} \bibinfo{author}{\bibfnamefont{J.~J.}
  \bibnamefont{Lacapere}}, \bibinfo{journal}{Biophys. J.}
  \textbf{\bibinfo{volume}{88}}, \bibinfo{pages}{818} (\bibinfo{year}{2005}).

\bibitem[{\citenamefont{Ikeguchi et~al.}(2005)\citenamefont{Ikeguchi, Ueno,
  Sato, and Kidera}}]{Kidera:05}
\bibinfo{author}{\bibfnamefont{M.}~\bibnamefont{Ikeguchi}},
  \bibinfo{author}{\bibfnamefont{J.}~\bibnamefont{Ueno}},
  \bibinfo{author}{\bibfnamefont{M.}~\bibnamefont{Sato}}, \bibnamefont{and}
  \bibinfo{author}{\bibfnamefont{A.}~\bibnamefont{Kidera}},
  \bibinfo{journal}{Phys. Rev. letters} \textbf{\bibinfo{volume}{94}},
  \bibinfo{pages}{078102} (\bibinfo{year}{2005}).

\bibitem[{\citenamefont{Zheng and Brooks}(2005)}]{Brooks:05}
\bibinfo{author}{\bibfnamefont{W.}~\bibnamefont{Zheng}} \bibnamefont{and}
  \bibinfo{author}{\bibfnamefont{B.~R.} \bibnamefont{Brooks}},
  \bibinfo{journal}{Biophys. J.} \textbf{\bibinfo{volume}{88}},
  \bibinfo{pages}{3109} (\bibinfo{year}{2005}).

\bibitem[{\citenamefont{Tirion}(1996)}]{Tirion:96}
\bibinfo{author}{\bibfnamefont{M.}~\bibnamefont{Tirion}},
  \bibinfo{journal}{Phys. Rev. Lett.} \textbf{\bibinfo{volume}{77}},
  \bibinfo{pages}{1905} (\bibinfo{year}{1996}).

\bibitem[{\citenamefont{Bahar et~al.}(1997)\citenamefont{Bahar, Atilgan, and
  Erman}}]{Bahar:97}
\bibinfo{author}{\bibfnamefont{I.}~\bibnamefont{Bahar}},
  \bibinfo{author}{\bibfnamefont{A.~R.} \bibnamefont{Atilgan}},
  \bibnamefont{and} \bibinfo{author}{\bibfnamefont{B.}~\bibnamefont{Erman}},
  \bibinfo{journal}{Folding \& Design} \textbf{\bibinfo{volume}{2}},
  \bibinfo{pages}{173} (\bibinfo{year}{1997}).

\bibitem[{\citenamefont{Hinsen}(1998)}]{Hinsen:98}
\bibinfo{author}{\bibfnamefont{K.}~\bibnamefont{Hinsen}},
  \bibinfo{journal}{Proteins} \textbf{\bibinfo{volume}{33}},
  \bibinfo{pages}{417} (\bibinfo{year}{1998}).

\bibitem[{\citenamefont{Brooks et~al.}(1983)\citenamefont{Brooks, Bruccoleri,
  Olafson, States, Swaminathan, and Karplus}}]{Charmm}
\bibinfo{author}{\bibfnamefont{B.~R.} \bibnamefont{Brooks}},
  \bibinfo{author}{\bibfnamefont{R.~E.} \bibnamefont{Bruccoleri}},
  \bibinfo{author}{\bibfnamefont{B.~D.} \bibnamefont{Olafson}},
  \bibinfo{author}{\bibfnamefont{D.~J.} \bibnamefont{States}},
  \bibinfo{author}{\bibfnamefont{S.}~\bibnamefont{Swaminathan}},
  \bibnamefont{and} \bibinfo{author}{\bibfnamefont{M.}~\bibnamefont{Karplus}},
  \bibinfo{journal}{J. Comp. Chem.} \textbf{\bibinfo{volume}{4}},
  \bibinfo{pages}{187} (\bibinfo{year}{1983}).

\bibitem[{\citenamefont{Lazaridis}(2003)}]{EEF11}
\bibinfo{author}{\bibfnamefont{T.}~\bibnamefont{Lazaridis}},
  \bibinfo{journal}{Proteins} \textbf{\bibinfo{volume}{52}},
  \bibinfo{pages}{176} (\bibinfo{year}{2003}).

\bibitem[{\citenamefont{Li and Cui}(2004)}]{Cui:04}
\bibinfo{author}{\bibfnamefont{G.}~\bibnamefont{Li}} \bibnamefont{and}
  \bibinfo{author}{\bibfnamefont{Q.}~\bibnamefont{Cui}},
  \bibinfo{journal}{Biophys. J.} \textbf{\bibinfo{volume}{86}},
  \bibinfo{pages}{743} (\bibinfo{year}{2004}).

\bibitem[{\citenamefont{Bruschweiler}(1995)}]{Local}
\bibinfo{author}{\bibfnamefont{R.}~\bibnamefont{Bruschweiler}},
  \bibinfo{journal}{J. Chem. Phys.} \textbf{\bibinfo{volume}{102}},
  \bibinfo{pages}{3396} (\bibinfo{year}{1995}).

\bibitem[{\citenamefont{Atilgan et~al.}(2001)\citenamefont{Atilgan, Durell,
  Jernigan, Demirel, Keskin, and Bahar}}]{Bahar:01}
\bibinfo{author}{\bibfnamefont{A.}~\bibnamefont{Atilgan}},
  \bibinfo{author}{\bibfnamefont{S.}~\bibnamefont{Durell}},
  \bibinfo{author}{\bibfnamefont{R.}~\bibnamefont{Jernigan}},
  \bibinfo{author}{\bibfnamefont{M.}~\bibnamefont{Demirel}},
  \bibinfo{author}{\bibfnamefont{O.}~\bibnamefont{Keskin}}, \bibnamefont{and}
  \bibinfo{author}{\bibfnamefont{I.}~\bibnamefont{Bahar}},
  \bibinfo{journal}{Biophys. J.} \textbf{\bibinfo{volume}{80}},
  \bibinfo{pages}{505} (\bibinfo{year}{2001}).

\bibitem[{\citenamefont{Goldstein}(1950)}]{Goldstein:50}
\bibinfo{author}{\bibfnamefont{H.}~\bibnamefont{Goldstein}},
  \emph{\bibinfo{title}{Classical Mechanics}}
  (\bibinfo{publisher}{Addison-Wesley}, \bibinfo{year}{1950}),
  \bibinfo{edition}{reading, ma} ed.

\bibitem[{\citenamefont{Kraulis}(1991)}]{Molscript}
\bibinfo{author}{\bibfnamefont{P.}~\bibnamefont{Kraulis}}, \bibinfo{journal}{J.
  Appl. Cryst.} \textbf{\bibinfo{volume}{24}}, \bibinfo{pages}{946}
  (\bibinfo{year}{1991}).

\end{thebibliography}
\end{document}